\documentclass[twocolumn,tightenlines,prd,nofootinbib,floatfix,superscriptaddress]{revtex4}

\usepackage{epsfig} 
\usepackage{amssymb} 
\usepackage{amsmath} 
\usepackage{amsfonts}

\newcommand{\N}{{\cal N}}

\newcommand{\be}{\begin{equation}}
\newcommand{\ee}{\end{equation}}

\newcommand{\nn}{\nonumber}

\newcommand{\Eref}[1]{Eq.~(\ref{#1})}

\newcommand{\gev}{{\rm GeV}}

\newcommand{\f}{\frac} 
\newcommand{\D}[2]{\frac{d{#1}}{d{#2}}}

\renewcommand{\vec}[1]{\mbox{\boldmath $#1$}}

\newcommand{\xgg}{\chi^{\prime\prime}}
\newcommand{\x}{\chi}

\newcommand{\gc}{\gamma_s}
\newcommand{\as}{\bar\alpha}
\newcommand{\als}{\alpha_s}

\begin{document}

\title{Prompt neutrino fluxes from atmospheric charm}

\author{Rikard Enberg}
\affiliation{Department of Physics, University of Arizona, Tucson, AZ 85721}

\author{Mary Hall Reno}
\affiliation{Department of Physics and Astronomy, University of Iowa, Iowa City, IA}

\author{Ina Sarcevic}
\affiliation{Department of Physics, University of Arizona, Tucson, AZ 85721}

\begin{abstract}
We calculate the prompt neutrino flux from atmospheric charm production by
cosmic rays, using the dipole picture in a
perturbative QCD framework, which incorporates the parton saturation effects 
present at high energies.  We compare 
our results with the next-to-leading order perturbative 
QCD result and find that saturation effects are large for neutrino 
energies above $10^6$ GeV, leading to a substantial suppression of the prompt 
neutrino flux.  We comment on the range of prompt neutrino fluxes due to 
theoretical uncertainties.  

\end{abstract}

\maketitle

\section{Introduction} 

Atmospheric neutrinos are produced in interactions of cosmic rays with 
Earth's atmosphere.  The observation of low energy ($E_\nu \sim$ GeV) 
atmospheric neutrinos, their flavor-dependent interactions, and their path
length dependence~\cite{SuperK, SNO} has confirmed the existence of neutrino
flavor transformation, and therefore the most fundamental property of the 
neutrinos:\ that they are not massless.  These observations 
 have provided a remarkable source of
information about mass and mixing parameters of neutrinos. 

Atmospheric neutrinos are also a
background to other sources of neutrinos, such as cosmogenic neutrinos produced in 
interactions of cosmic rays with
the background radiation ~\cite{ess} and directly
from sources such as 
active galactic nuclei and gamma ray bursts~\cite{rev}. Observation
of neutrinos coming from these distant sources 
would provide valuable information 
about the particle production mechanism in 
astrophysical sources.  

Neutrino interactions in the Earth and in the atmosphere
could also serve as unique probes of physics beyond the 
Standard Model \cite{nonstd} .
It has been recently suggested that atmospheric neutrinos, in their
interactions in the Earth, could produce supersymmetric particles if the
neutrino energies are sufficiently high.  In Ref.~\cite{ando},
Ando et al.\ have suggested that the high energy atmospheric neutrino flux
may be large enough to produce quasi-stable charged particles that are
potentially detectable in the IceCube~\cite{icecube}
neutrino detector.
As a background to high energy sources or as a flux to produce
exotic particles in the Earth, it is useful to re-evaluate the
high energy component of the atmospheric neutrino flux.
  
The atmospheric fluxes of neutrinos at low energies have been extensively
studied~\cite{agrawal,battistoni,honda,gaisserhonda}. 
They arise mainly from the products of charged pion
and kaon decays. As energies increase, the decay lengths of the mesons become
longer than their path lengths in the atmosphere~\cite{pathlength},
suppressing the production of neutrinos. Other, shorter lived hadrons are also
produced at high energies. They too contribute to the neutrino flux, especially from the 
``prompt'' decay of charmed mesons.  
The energy dependence of these prompt neutrinos is less steep than the 
``conventional'' neutrino flux from pion and kaon decays.  
The energy at which the prompt neutrinos become the dominant atmospheric
neutrino component depends on the 
details of the mechanism for charm production in proton--air collisions 
at high energies. Charm contributions to the atmospheric lepton fluxes
have been evaluated analytically and semi-analytically 
\cite{volkova,Gondolo:1995fq,Pasquali:1998ji,graciela,Martin:2003us}.
There are renewed efforts include charm production with the
dual parton model in Monte Carlo
simulations of air showers as well \cite{bat,Berghaus:2007hp}.

For vertical neutrino fluxes, the cross-over between 
conventional and prompt dominated fluxes occurs in the
energy range of $10^5$--$10^6$ GeV for the calculations of
Refs. \cite{Gondolo:1995fq,Pasquali:1998ji,graciela,Martin:2003us},
and the cross-over energy increases with zenith angle.
For energies above $\sim 10^6$~GeV, the dominant contribution to charm production 
comes from gluons, where saturation effects~\cite{glr} due to dense, 
interacting gluons in the nucleus become important.  
We evaluate the prompt neutrino flux using perturbative QCD in the dipole 
framework, taking these effects into account.  We study the theoretical 
uncertainties inherent in this approach and compare with standard next-to-leading 
order perturbative QCD. The range of QCD-based predictions yields prompt neutrino
fluxes that are unlikely to be large enough to produce a detectable number of
exotic particles of the type discussed in Ref.~\cite{ando}.

We begin with a discussion of the cross section for charm production in the
dipole picture in Section II. Using the dipole picture results, we discuss the
evaluation the prompt neutrino
flux from charm decays in Section III. Our results and a comparison with the 
conventional fluxes are shown in Section IV. We also discuss uncertainties in
the QCD approach, and compare our results with the uncertainty band of
Ref.~\cite{ando} in Section IV.

\section{Cross section for charm production}

\subsection{Charm production in perturbative QCD} 
 
In the perturbative QCD approach, 
the dominant contribution to the charm cross section at high energies comes from 
the partonic subprocess $gg\rightarrow c\bar{c}$. 
The parton-level differential cross section for production of 
$c\bar{c}$ pairs in 
proton--proton collision, 
at the leading order in the strong 
coupling constant, $\alpha_s(\mu^2)$, 
is given by 
\begin{equation}
\frac{d\sigma_{\rm LO}}{dx_F}=\int
\frac{ dM_{c\bar{c}}^2}{(x_1+x_2) s}
\sigma_{gg\rightarrow c\bar{c}}(\hat{s}) G(x_1,\mu^2) G(x_2,\mu^2)
\end{equation}
where $x_{1,2}$ are the momentum fractions of the gluons, $x_F=x_1-x_2$ is the Feynman variable,  
$G(x,\mu^2)$ is the gluon distribution of the proton, and $\mu$ is the factorization scale. 
Given the charm--anticharm invariant mass $M_{c\bar{c}}$, the 
fractional momenta of the gluons, $x_{1,2}$, can be expressed in terms of the 
the Feynman variable, $x_F$,  
\be
\label{eq:x12}
x_{1,2} = \frac{1}{2}\left( \sqrt{x_F^2+\frac{4M_{c\bar c}^2}{s}} 
\pm x_F\right) \ .  
\ee
Typically the factorization scale is taken to be of the order of $2m_c$.  

For the flux calculation we need the differential cross section
as a function of incident proton energy ($E_p$) and final charm energy
($E_c$), convoluted
with the incident cosmic ray proton flux.  
Clearly at high energies, given the relationship of Eq.\ (\ref{eq:x12}),
the charm cross section has dominant contribution when one gluon parton distribution 
function (PDF) is at $x_1 \sim x_F$
and the other gluon distribution is at $x_2\ll 1$.  
Since the gluon distribution cannot be measured directly, its value at 
 very small $x$ has large uncertainties, especially
for the low factorization scale $\mu\sim 2 m_c$.  The dipole picture gives a theoretically 
motivated description of small $x$ physics which can effectively take into account 
resummation of the large $\alpha_s\ln(1/x)$ contributions \cite{bfkl} to the evolution
of the PDFs.  
Thus by using the dipole picture, we avoid the large
uncertainty due to the unknown behavior of the gluon distribution at very small $x$.

The dipole picture is most straightforwardly described in the DIS context, which
we do next. We then elaborate how this is applied to hadron--hadron scattering.

\subsection{Dipole picture formalism in deep inelastic scattering} 

In deep inelastic lepton--hadron scattering, 
the high $Q^2$ virtual photon can penetrate the
nucleon and probe the partonic degrees of freedom. This 
partonic interpretation based on perturbative QCD is most relevant
in the infinite momentum frame.
The $Q^2$-dependence 
of the nucleon structure function $F_2^N(x,Q^2)$ is well accounted
for by the DGLAP evolution equations 
\cite{glap} given some non-perturbative initial condition $F_2^N(x,Q_0^2)$. 
As noted above, at small $x$ one needs to consider 
the resummation of large logarithms $\ln 1/x$, which leads 
to the BFKL evolution equation \cite{bfkl}.  

Another feature of the nucleon structure function $F_2^N$ in the DGLAP
framework is the strong growth of the gluon
density in the nucleon in the small $x$ region. 
In the infinite momentum frame, because of
the high nucleon and parton densities, quarks and gluons that belong
to different nucleons in the nucleus may recombine and annihilate, leading to
the recombination effect first proposed by Gribov, Levin and
Ryskin (GLR)~\cite{glr} and later detailed by Mueller and Qiu~\cite{mq}. 

An alternative approach is to consider instead the interaction in the
target rest frame (laboratory frame), where the virtual photon interacts with 
nucleons via its quark--antiquark pair ($q\bar q$) color-singlet 
fluctuation \cite{lu}. If the coherence
length of the virtual photon fluctuation is larger than the radius of the 
nucleus, $l_c>R_A$, the $q\bar q$ configuration interacts
coherently with all nucleons, with a cross section given by the
color transparency mechanism for a pointlike color-singlet configuration
\cite{plc}. That is, the cross section is proportional to the transverse
separation squared, $r^2$, of the $q$ and $\bar q$. 

\begin{figure}
\centerline{\epsfig{file=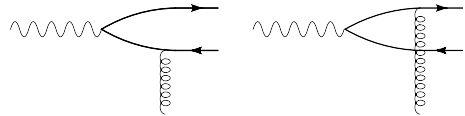,width=0.7\columnwidth}}
\caption{The perturbative diagrams giving rise to scattering with a
gluon of the
$\gamma^* \to q\bar q$ fluctuation in deep inelastic scattering.}
\label{fig:1}
\end{figure}

In the dipole picture, the cross
section for the absorption of a virtual photon in the small $x$ region
is dominated by the scattering of a gluon off the $q\bar q$ pair fluctuation
of the virtual photon. 
The
generic perturbative QCD diagrams giving rise to the $q\bar q$ fluctuation
are shown in Figure~\ref{fig:1}. 
The invariant mass  of the incoming virtual
photon-proton system at small $x$ is related to the photon virtuality $Q^2$ by
\begin{eqnarray}
s=(q+p)^2\simeq 2p\cdot q=\frac{Q^2}{x}\; ,
\end{eqnarray}
where $q$ and $p$ are the four-momenta of the photon and the target nucleon, $q^2=-Q^2$ and $x=Q^2/2p\cdot q$.
Thus the region of small $x$ corresponds to a high energy scattering process 
at fixed $Q^2$.

The imaginary part of the sum of the amplitudes in Figure \ref{fig:1} is related
to the photoabsorption cross section, which has been calculated by
Nikolaev and Zakharov \cite{Nikolaev:1990ja} assuming that the size of the  
$q\bar q$ pair is frozen in the scattering process and that the one-gluon 
exchange process of Figure \ref{fig:1} dominates. 
The transverse cross section can be cast into an impact 
parameter representation 
\begin{eqnarray}
\sigma (\gamma^*N)=\int_0^1dz \int d^2{\bf r}|\Psi_T (z,{\bf r},Q^2)|^2
\sigma_{q\bar qN}(x,{\bf r})\; , \label{eq:sig}
\end{eqnarray}
where $z$ is the Sudakov variable, defined to be the fraction of 
the $q\bar q$ pair momentum carried by the quark, and
${\bf r}$ is the variable conjugate to the transverse momentum
of  the quark, representing the transverse size of the
pair. 
The function
$\Psi_T (z,{\bf r},Q^2)$ can be interpreted as the wave function
of the $q\bar q$ fluctuations of the virtual photon. Thus, 
$|\Psi_T (z,{\bf r},Q^2)|^2$ is the probability of finding 
a $q\bar q$ pair with a separation ${\bf r}$ and a fractional
momentum $z$. It is given for each quark flavor $f$ with
fractional charge $e_f$ by \cite{Nikolaev:1990ja}
\begin{align}
&|\Psi^f_{T}(z,\vec r,Q^2)|^2 = \\
&e_f^2 \frac{\alpha_{em} N_c}{2\pi^2} 
\left[ \left( z^2 + (1-z)^2 \right)
\epsilon^2 K_1^2(\epsilon r) + m_f^2 K_0^2(\epsilon
r)\right],\nn
\end{align}
where $\epsilon^2 = z(1-z)Q^2 +  m_f^2$, and $K_0$
and $K_1$ are modified Bessel functions.  

The cross section for the high energy interaction of a small-size $q\bar q$ configuration
with the nucleon, $\sigma_{q\bar qN}({\bf r})$, can be calculated in 
leading-order perturbative QCD. In this approximation,
one sets $\sigma_{q\bar{q}N}(\vec{r})$ equal to~\cite{fixedorder}
\be
\sigma_d^{pQCD} = \f{\pi^3}{3} \, r^2 \, \alpha_s(\mu) \, x \, G(x_1,\mu^2).
\ee 
 This cross section is, as discussed above, proportional to the
square of the size of the pointlike configuration as a consequence of color
transparency in QCD.  
However, 
the singular behavior of the
wave function and the strong
scaling violation of the gluon distribution in the small-$x$ region as 
$r$ decreases can compensate the smallness of the cross section due to
color transparency. 

Ultimately, gluon saturation effects need to be included
for a more realistic $\sigma_{q\bar{q}N}(\vec{r})$. One would then 
 derive an approximate expression for the dipole cross section 
from theory, \emph{including} saturation effects, and use experimental data to 
determine incalculable parameters in this expression.
Before we
turn to saturation and the types of functional forms used to fit the
dipole cross section, 
in the next section we describe how heavy quark production in
proton-proton scattering is treated in the dipole picture.

\subsection{Heavy quark production} 

\begin{figure}
\centerline{\epsfig{file=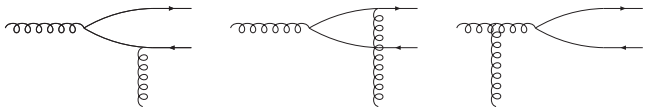,width=\columnwidth}}
\caption{The perturbative diagrams giving rise to the scattering of a
gluon with 
the $g\to q\bar q$ pair fluctuation in hadronic collisions.}
\label{fig:2}
\end{figure}

Heavy quark production in hadronic collisions can be obtained in the same 
formalism 
\cite{Nikolaev:1995ty,Raufeisen:2002ka,Kopeliovich:2002yv,Goncalves:2006ch}.  
In this case, the dipole is produced from a gluon instead of a photon, so that 
the dipole can be in
a color octet state. As shown in Figure \ref{fig:2}, there is now an additional 
diagram that contributes, in
which the gluon interacts with the target before fluctuating to a dipole.  

The differential cross
section for heavy quark production is \cite{Nikolaev:1995ty}
\begin{align}
\frac{d\sigma(pp\to Q\bar
Q X)}{dy} \simeq  
x_1 \, G(x_1,\mu^2) \, \sigma^{G p \to Q\bar Q
X}(x_2,\mu^2,Q^2),
\label{dsigmady}
\end{align}
where $x_1$ and $x_2$ are the
partonic momentum fractions, 
$y=\frac{1}{2}\ln(x_1/x_2)$ is the $Q\bar Q$ pair rapidity and $\sigma^{G p
\to Q\bar Q X}$ is the partonic cross section calculated in the dipole
model,
\begin{align}
\sigma^{G p \to Q\bar Q X}(x,\mu^2,Q^2) = \int dz \, d^2\vec r 
|\Psi^Q_G(z,\vec r)|^2 \sigma_{dG}(x,\vec r)\ .
\end{align}
The probability of finding a $Q\bar Q$ pair with a separation $\vec r$ and a 
fractional momentum z, is given by 
\begin{align}
&|\Psi^Q_G(z,\vec r,Q^2=0)|^2 =  \\
&\frac{\alpha_s(\mu)}{2\pi^2} 
\left[ \left( z^2 + (1-z)^2 \right)
 m_Q^2 K_1^2(m_Q r) + m_Q^2 K_0^2(m_Q r)\right],\nn
\end{align}
where $\mu \sim 1/r$ is the factorization scale. For heavy quark production we
have $Q^2=0$, so $\mu\sim m_Q$ and $\epsilon=m_Q$.

The dipole cross section that describes the interaction of a heavy quark--antiquark
pair from the fluctuation of a gluon with the target nucleon is given by 
\cite{Nikolaev:1995ty} 
\begin{align}
\label{eq:sigdg}
\sigma_{GQ\bar Q}^N (x_2,\vec r) &= \frac{9}{8} \left[\sigma_d(x_2,z\vec r)
+\sigma_d(x_2,(1-z)\vec r)\right]  \nn\\
&-\frac{1}{8}\sigma_d(x_2,\vec r),  
\end{align}
where 
$\sigma_d$ 
is the color singlet dipole cross section of Eq. (4).  
 The first term corresponds to 
the quark--gluon ($G-Q$) separation $z\vec r$, the antiquark--gluon ($G-\bar Q$)
separation $(1-z)\vec r$ and the quark--antiquark ($Q-\bar Q$) separation 
$\vec r$.  This expression includes contributions from the 
three different color and spin states 
in which $Q\bar Q$ can be produced 
 \cite{Kopeliovich:2002yv}.

Finally, to take threshold corrections for charm production at large $x$ into
account, the dipole cross section is multiplied with a factor 
$(1-x_2)^7$~\cite{Motyka:2002ww}. We find this correction to be 
negligible for energies 
 above $10^3$~GeV.

\subsection{The dipole--proton cross section and saturation}

The dynamics of the scattering process at small $x$ is, in principle,
included in the 
dipole cross section. Thus, to compute the differential 
cross section 
$d\sigma/dx_F$ we must find the cross section for a $c\bar c$ dipole to 
scatter on the proton, including the effects of saturation.

A simple model for saturation was proposed by Golec-Biernat and 
W\"usthoff~\cite{GolecBiernat:1998js}.  In their model, the dipole 
cross section is parametrized as
\be
\label{eq:gbw}
\sigma_d^{GBW} = \sigma_0\left[1-e^{-r^2 Q_s^2(x)/4}\right],
\ee
where $Q_s$ is the saturation
scale, 
\be
\label{eq:qs}
Q_s=Q_s(x)=Q_0(x_0/x)^{\lambda/2}
\ee 
with $Q_0=1$~GeV. The parameters $\lambda$ and $x_0$ in the above expressions 
were fitted to HERA data on the structure function $F_2$ and the
diffractive structure function $F_2^D$~\cite{GolecBiernat:1998js}.

This is a phenomenological model, constructed to give the right
behavior of the dipole cross section in the two limits  $r\to 0$ and
$r\to\infty$. Eq.\ (\ref{eq:gbw}) has
$\sigma \propto r^2$ for small $r$, as implied by perturbative QCD, and 
$\sigma\to$~const for large $r$ (this is the saturation property of the cross
section), thus providing 
some insight into the physics of saturation.
The simple
parametrization also gave a good fit to the data, although
it does not reproduce 
newer data as well~\cite{Bartels:2002cj} as it does the older data. 

One would like to calculate the dipole cross section rigorously in perturbative
QCD; however, it is not known how to fully include the effects of saturation. 
It is convenient to study QCD evolution in Mueller's 
dipole formulation~\cite{Mueller:1993rr}, where the projectile contains a
collection of color dipoles. It has been shown~\cite{stochastic} that in 
the high-energy limit, the scattering process is equivalent to a 
stochastic reaction--diffusion process where there are
fluctuations in the number of dipoles. These fluctuations may potentially have a large effect on
the energy dependence of the amplitude and saturation scale. A full
calculation should include these effects, but 
 they were found to be small in the region of very small $x$~\cite{Enberg:2005cb}.  
In principle one should also 
take into account the complicated dynamics of the
color glass condensate \cite{MV,JIMWLK,Balitsky,Gelis:2008rw}. This is described by the functional integro-differential
Jalilian-Marian, Iancu, McLerran, Weigert, Leonidov and Kovner
(JIMWLK) equations~\cite{JIMWLK}, or 
equivalently by Balitsky's infinite 
hierarchy of coupled differential equations for the 
expectation values of Wilson lines~\cite{Balitsky}. 

A much simpler equation which includes saturation
was obtained by Balitsky~\cite{Balitsky} 
and Kovchegov~\cite{Kovchegov} in the particular case where the target is a 
large nucleus. This equation is known as the Balitsky--Kovchegov (BK) equation
and although it can be derived within the dipole framework, it turns out to
represent a specific mean-field approximation to the Balitsky--JIMWLK equations. 
 The BK equation is, like the BFKL equation, a leading logarithmic evolution 
equation in $\ln(1/x)$. The BFKL equation, however, is a linear equation, while
the BK equation is similar in structure to the GLR~\cite{glr} and
Muller--Qiu~\cite{mq} equations, and can be written as the BFKL
equation modified by a non-linear term. 
This reduces the power
growth of the gluon distribution as $x\to 0$, which has
been established by both numerical and approximate analytical
studies, see e.g.\ Ref.~\cite{Enberg:2005cb} and references therein.

The dipole cross section is obtained from the solution of the BK 
equation, which can be solved numerically.  We will 
 instead use an approximate 
result~\cite{Iancu:2003ge}, which consists of a matching of approximate 
analytic solutions of the BK equation in the two regions of dipole size 
$r \gg 1/Q_s$ and $r \ll 1/Q_s$, where the equation simplifies.

In the large $r$ region the solution approaches a fixed
saturated value as $r\to \infty$ \cite{Levin:1999mw}. For $r\ll 1/Q_s$ 
the effects of the non-linearity in the BK equation are small, and 
the equation reduces to the BFKL equation; 
the solution is a saddle point solution of the 
BFKL equation, subject to a saturation condition~\cite{Iancu:2003ge}.
The two solutions are matched at an intermediate scale $r Q_s=2$. 
The resulting model is what we will refer to as the dipole model (DM).
 
The dipole cross section is given by 
\be
\sigma_d(x,\vec r)=\sigma_0 \N (r Q_s, Y),
\label{sigmadip1}
\ee 
where $\sigma_0$ is a constant, and $\N$ is the 
forward dipole scattering amplitude 
 obtained from the BFKL or BK equation 
 \cite{Iancu:2003ge},
\begin{equation}
\N(r Q_s, Y) = 
\begin{cases}
\N_0
\left(\dfrac{\tau}{2}\right)^{2\gamma_\text{eff}(x,r)}, \qquad &\text{for }
\tau<2 \\
1-\exp\left[-a \ln^2 (b\tau)\right], & \text{for }
\tau>2\ .
\end{cases}
\label{sigmadip2}
\end{equation}
Here $\tau = r Q_s$, $Y=\ln(1/x)$ is the rapidity, and again the
saturation scale is defined in Eq.\ (\ref{eq:qs})
with $Q_0=1$~GeV.
Furthermore,
\be
\gamma_\text{eff}(x,r)=\gamma_s+\frac{\ln (2/\tau)}{\kappa \lambda
Y}
\ee
is the ``effective anomalous dimension,'' and $\gamma_s$ and $\kappa$ are
theoretical parameters calculated from the BFKL equation, 
with numerical values $\gamma_s=0.627$ and $\kappa=9.94$. Note that this is a 
perturbative QCD result and not an \emph{ad hoc} model, 
although it is obtained by an approximate solution of the BK equation.

The free parameters in the model are $\N_0$, $\sigma_0$,
$\lambda$ and $x_0$. In Ref.~\cite{Iancu:2003ge}, the first of these was chosen to take the value 0.7.
The exponent $\lambda$ specifies the power behavior of the saturation
scale with $x$, and $x_0$ is the value of $x$ where the saturation scale is 1
GeV. Furthermore, $a$ and $b$ are matching coefficients to be chosen such that
the dipole amplitude and its derivative with respect to $r$ are continuous 
at $\tau=2$. We find
\begin{align}
a &=-\frac{\ln (1-{\N_0})}{\ln ^2(1-{\N_0})^{\frac{1}{\gamma_s
   }-\frac{1}{{\N_0} \gamma_s }}}\\
b &= \frac{1}{2} (1-\N_0)^{\frac{1}{\gamma_s }-\frac{1}{\N_0 \gamma_s }}.
\end{align} 
Note that the amplitude is a function of $r$ and $x$ only in the
combination indicated, $\tau\equiv r Q_s(x)$, except for the geometric scaling
breaking term in the effective anomalous dimension which contains the rapidity.

\begin{table}
\begin{center}
\begin{tabular}{lccccc}
\hline\hline
Ref. & $\N_0$ & $\gamma_s$ & $\lambda$ & $x_0$ & $\sigma_0$ (mb)   \\
\hline
\protect\cite{Iancu:2003ge} 
&0.7
&0.627
& $0.253$ 
& $0.267\times 10^{-4}$ 
& $25.8$ \\
\cite{Stephane2}
&$0.7$
&0.627
&$0.175$
&$0.19 \times 10^{-6}$
&$37.3$ \\ 
\cite{Soyez:2007kg}
&$0.7$
&0.738
&$0.220$
&$0.163 \times 10^{-4}$ 
&$27.3$\\
\hline\hline
\end{tabular}
\caption{Parameter values in the dipole cross section formulas
in Eqs.\ (\ref{sigmadip1}) and (\ref{sigmadip2}). In Refs.\
\protect\cite{Iancu:2003ge} and \protect\cite{Stephane2}, $\gamma_s$ is calculated,
while in Ref.\ \protect\cite{Soyez:2007kg}, it is a fit parameter.}
\label{table-parameters}
\end{center}
\end{table}

The fitted parameter values from three different fits to HERA data on the 
deep inelastic structure function $F_2$ at small
$x$~\cite{Iancu:2003ge,Stephane2,Soyez:2007kg} are shown in 
Table~\ref{table-parameters}. Note that in all cases $\N_0$ was fixed at 
$\N_0=0.7$. The first row shows the original parameter values obtained 
in Ref.~\cite{Iancu:2003ge}. This was a three-flavor fit and is 
therefore not suitable for our calculation, but it has been extended 
to include charm~\cite{Stephane2}, giving the values in the second row. 
Finally, the third row shows a more recent fit by Soyez~\cite{Soyez:2007kg}, 
which also includes charm. In this fit the parameter
$\gamma_s$ was taken as a free parameter, which gave a better fit to the data,
with a larger value of $\gamma_s$ and a smaller value of $\lambda$. 
This is quite interesting since a reduction of these parameters is exactly what
is expected when including higher order logarithmic corrections to the BFKL
kernel in the BK equation~\cite{NLL}. 

These models take into account only the leading
exponential $x$-dependence of the saturation scale, and there are large
sub-asymptotic corrections to the energy 
dependence~\cite{MunierPeschanski},
\begin{align}
\ln Q^2_s(Y) &= \frac{3\als}{\pi} \frac{\x(\gc)}{\gc} Y 
- \frac{3}{2\gc} \ln Y \nn \\
&-\frac{3}{\gc^2} \sqrt{\frac{2\pi}{\as\xgg(\gc)}} \frac{1}{\sqrt{Y}}
+{\cal
O}(1/Y),
\label{Qs-LL-fixed}
\end{align}
where $\chi(\gamma)$ is the BFKL
characteristic function and $\gc=0.627$. The models discussed in this
section keep only the leading term in this expression. Using the full
expression could potentially change the energy dependence of the cross section
substantially, but to incorporate this in this dipole model would require
introducing more parameters and performing a new fit to all the data.

In the dipole model results that
follow, the DM results shown use the parameters 
of Soyez \cite{Soyez:2007kg}
shown in Table \ref{table-parameters} and the parametrization of
equations (\ref{sigmadip1}--\ref{sigmadip2}).

\subsection{Nuclear effects}

In a dipole framework there are two 
 possible ways to include nuclear effects suggested in the literature:\ 
modification of 
the saturation scale, e.g.\ as proposed by Armesto, Salgado and
Wiedemann (ASW)~\cite{Armesto:2004ud} 
(see also \cite{Kowalski} for another approach), and 
the Glauber--Gribov~\cite{Glauber,Gribov:1968jf} formalism.  
In the former case, the nuclear effects are
accounted for by geometric scaling, simply scaling the saturation scale for a
nucleus $A$ according to
\be
Q_{s,A}^2 = Q_{s,p}^2
\left( \dfrac{A\pi R_p^2}{\pi R_A^2} \right)^{1/\delta}
\label{QsA}
\ee
where $R_p$ is the proton radius, $R_A=1.12 A^{1/3}-0.86 A^{-1/3}$~fm is the
nuclear radius, and $\delta$ is a free parameter to be fitted to data. ASW find
$\delta=0.79$ by fitting to $\gamma^* A$ data at small $x$. The proton
radius is related to $\sigma_0$ in the dipole cross section through
$\sigma_0=2\pi R_p^2$. 

In the Glauber--Gribov formalism, nuclear rescattering is taken into account by integrating the 
 dipole cross section for dipole--nucleus collisions over the
impact parameter,
\be
\sigma_d^A(x,r) = \int d^2b\sigma_d^A(x,r,b),
\ee
where $b$ is the impact parameter between the center of the dipole and the
center of the nucleus. The expression for the $b$-dependent cross section is
given by 
\be
\sigma_d^A(x,r,b) = 2\left[1- \exp\left(-\frac{1}{2}A \, T_A(b)\, 
\sigma_d^p(x,r)\right)\right],
\ee
where $\sigma_d^p(x,r)$ is the dipole--proton cross section given in Eqs.\
(\ref{sigmadip1}) and (\ref{sigmadip2}) and $T_A(b)$ is the nuclear profile
function,
\be
T_A(b) = \int dz \, \rho_A(z,\vec b),
\ee
where $\rho_A$ is the nuclear density, and $T_A$ is normalized so that
\be
\int d^2b \, T_A(b) = 1. 
\ee
This model has e.g.\ been used in Ref.\ \cite{Armesto:2002ny} with a Fermi 
distribution for $\rho$ to compute nuclear structure functions with 
good results. 

We compared the Glauber--Gribov model with a Gaussian distribution for 
$\rho$ to the ASW method and found that for the relatively light air nuclei, these
two methods give very similar results (within 10\%). We will in the following 
use the simpler ASW method.

Predictions from the framework described above have been tested against data.
For deep inelastic structure functions this was done 
in Refs.~\cite{Iancu:2003ge,Soyez:2007kg}. The ratio of DIS on nuclei to DIS on
deuterons was calculated and compared to E665 data in Ref.~\cite{Cazaroto:2008iy},
and total cross sections for $\gamma p$, $\gamma A$, $p p$, and $p A$ were
calculated in~\cite{Goncalves:2006ch} using the parameters from
\cite{Stephane2} (second row in Table~\ref{table-parameters}), and the $\gamma
p$ and $p p$ results were compared to data with good agreement. 
There have
been no tests in the energy range probed by cosmic rays. However, the LHC
will begin to access these energy scales shortly.

\subsection{Fragmentation of charm quarks} 

Our earlier analytical calculation \cite{Pasquali:1998ji} did not take 
fragmentation of the charm quarks into charmed hadrons into account, but simply
took the hadron to have the same energy as the charm quark. In Ref.
\cite{Martin:2003us}, fragmentation was taken into account by decreasing the
momentum fraction of the hadron to an average lower value. In this paper we
take fragmentation into account by using fragmentation functions. 

For our comparison without fragmentation, we
use updated hadron fractions~\cite{Yao:2006px} 
\be
\label{eq:ff}
f_{D^0}=0.565,\
f_{D^+}=0.246,\ 
f_{D_s^+}=0.080, \ 
f_{\Lambda_c}=0.094
\ee
where
$f_h$ is the fraction of fragmentation of $c\to h$. These newer values are somewhat
different from the values used 
in~\cite{Gondolo:1995fq,Pasquali:1998ji,Martin:2003us}; this increases the
computed flux by about 20\%.

In general the cross section for hadron production 
including fragmentation is obtained from the cross
section for charm production as
\be
\D{\sigma(p p\to h X)}{E_h} = \int_{E_h}^\infty \frac{dE_c}{E_c}
\D{\sigma(p p\to c X)}{E_c} D_c^h(E_h/E_c), 
\ee
where $D_c^h(z)$ is the fragmentation function for $c\to h$. This can be
written
in terms of momentum fractions as
\be
\D{\sigma(p p\to h X)}{x_E} = \int_{x_E}^1 \frac{dz}{z}
\D{\sigma(p p\to c X)}{x_c} D_c^h(z), 
\ee
where $z=E_h/E_c$, $x_c=E_c/E_p$, and $x_E=E_h/E_p$. At high energy the
momentum
fraction $x_E\simeq x_F$, or, for the charm cross section, $x_c\simeq x_F$.

We use both the older Peterson fragmentation function \cite{Peterson:1982ak}
and
the recent parametrization by Kniehl and Kramer (KK) in 
Ref.~\cite{KniehlKramer}. The Peterson function is given by
\be
D_c^h(z)=N_h\frac{1}{z}\left(1-\frac{1}{z}-\frac{\epsilon }{1-z}\right)^{-2},
\ee
where $\epsilon=0.05$ is a fitted parameter \cite{Yao:2006px}, common for all
mesons, and $N_h=f_h N$ is a
normalization constant where $N$ is given by the condition
\be
\sum_h \int dz D_c^h(z) = 1,
\ee 
assuming that the shape of $D_c^h$ is independent of the hadron $h$, 
and the fragmentation fractions
are given in Eq.\ (\ref{eq:ff}). The calculation without fragmentation 
amounts to taking fragmentation functions $D_c^h(z)=f_h \delta(1-z)$.

The KK fragmentation function has the form
\be
D_c^h = N_h \frac{x(1-x)^2}{\left[(1-x)^2+\epsilon_h x\right]^2},
\ee
with the parameters given in Ref.\ \cite{KniehlKramer}, which we show
in Table \ref{table:kk}.

\begin{table}
\centering
\begin{tabular}{c c c}
\hline\hline
Hadron $h$ & $N_h$ & $\epsilon_h$\\ [0.5ex]
\hline
$D^0$ & 0.694 & 0.101 \\
$D^+$ & 0.282 & 0.104 \\
$D^+_s$ & 0.050 & 0.032\\
$\Lambda_c^+$ & 0.00677 & 0.00418\\
[1ex]
\hline
\end{tabular}
\caption{Parameters in the LO Kniehl and Kramer fragmentation model
\cite{KniehlKramer}.}
\label{table:kk}
\end{table}
The Kniehl--Kramer fragmentation
functions have normalization factors fitted to the data. The integrals of these
functions give the fragmentation fractions in the KK model, and these are quite
different from the values cited above:\ for the LO fit we obtain
$f_{D^0}=0.745$, 
$f_{D^+}=0.296$,
$f_{D_s^+}=0.125$, and
$f_{\Lambda_c}=0.063$.
KK also perform a NLO fit. The NLO values decreases the normalization of the
calculated flux by about 10\%, but as our calculation is a LO calculation it is
more consistent to use the LO fit. These fragmentation fractions do not add to one,
an indication of one of the theoretical uncertainties.

\subsection{Theoretical uncertainties in the charm pair
cross section} 

Because the charm quark mass is of order 1 GeV, there are in principle large
uncertainties in
the charm pair production cross section \cite{Vogt:2007aw}. In
perturbation theory using
parton distribution functions, the charm cross
section predictions can vary by more than an
order of magnitude depending on
the charm quark mass, number of flavors and choice of scales and PDFs.
The
dipole approach, with the fit to DIS data then translated to hadron scattering,
mitigates the
uncertainty. Beyond the total cross section, one is interested in
the energy distribution of the
charmed quark.

\begin{figure} 
\begin{center} 
\epsfig{file=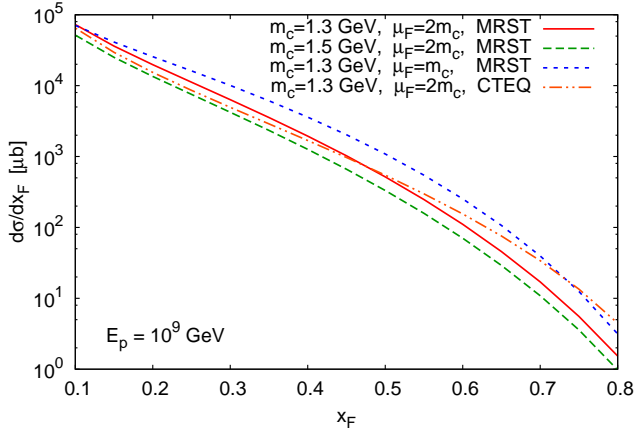, width=\columnwidth}
\caption{Charm quark $x_F$ distribution in proton--air collisions at 
$E_p=10^9$~GeV, calculated in the  dipole model described in the text,
with the ``standard'' choices of $m_c=1.3$ GeV, factorization scale
$\mu_F=2m_c$ and the MRST2001 PDFs \cite{mrst}.
\label{fig:cgc}}
\end{center}
\end{figure}

To investigate the sensitivity of the charm differential cross section to the 
choice of parameters, we vary them as follows:\ We use the parameters of 
Ref.\ \cite{Soyez:2007kg} for the dipole cross section (the fit of 
Ref.\ \cite{Stephane2} gives very similar results). We vary the PDF by taking 
the MRST 2001 LO \cite{mrst} or the CTEQ 6L gluon distributions \cite{cteq}, 
and we vary the factorization scale between $\mu_F=2 m_c$ or $\mu_F=m_c$, where
the charm quark mass is varied between $m_c=1.3$~GeV and $m_c=1.5$~GeV. 
In each of the listed cases the former choice
is what we use as our ``standard'' curves below.  In Figure~\ref{fig:cgc} we 
show a representative set of predictions for the differential cross section
$d\sigma(pA\to c\bar c)/dx_F$ for $A=14.5$, the average nucleon
number of air, and an incident proton energy of 
$10^9$~GeV. The parameter combinations that are not shown in the plot give 
results that fall between the upper and lower lines.

\begin{figure} 
\begin{center} 
\epsfig{file=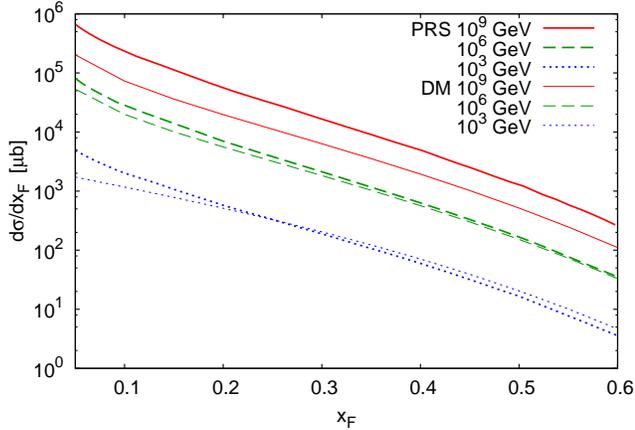, width=\columnwidth}
\caption{The NLO QCD $pA\rightarrow c\bar{c}X$ differential
cross section as a function of Feynman $x_F$ 
for PRS \cite{Pasquali:1998ji} compared to the dipole model
(DM) result for incident proton energies of
$10^3,\ 10^6,\ 10^9$ GeV. The thicker lines are PRS and the thinner lines with the same
color are DM at the same energy.
\label{fig:prscgc}}
\end{center}
\end{figure}

We are also interested in the difference between the predictions of NLO QCD and
the saturation prediction of the DM model. This is illustrated in
Figure~\ref{fig:prscgc}, where we show $d\sigma(pA\to c\bar c)/dx_F$ at three
energies using these two calculations.
The NLO QCD cross section come from
Ref.\ \cite{Pasquali:1998ji} (PRS). Note that the NLO QCD cross section
increases with energy much faster than the DM cross section. For the lower
energy $E=10^3$~GeV the cross sections are comparable, but we shall see that
because of the different energy dependence, the neutrino flux calculated using
NLO QCD is larger than the one calculated from the DM model.

\section{Calculation of neutrino fluxes}

The lepton flux at sea level is calculated by solving the coupled set of
differential equations that describes the cascade in the atmosphere initiated
by the incident cosmic ray nucleons. We use the primary nucleon flux
parametrization with a knee from Ref.
\cite{Gondolo:1995fq}:
\begin{align}
\phi_N(E)= 
\begin{cases}
1.7 \, E^{-2.7}  \quad  &\text{for } E<5 \cdot 10^6 \,\, \gev \\ 
174 \, E^{-3}           &\text{for } E>5 \cdot 10^6 \,\, \gev,  
\end{cases}
\end{align}
where the cosmic ray energy $E$ is given in \gev{} and the flux $\phi_N(E)$ in 
$\text{cm}^{-2} \, \text{s}^{-1} \, \text{sr}^{-1} \, (\gev/A)^{-1}$.   

The cascade consists of production and
attenuation through interactions and decay of the particles. We follow the
analytic approximation method used in Refs.
\cite{GaisserBook,Lipari:1993hd,Gondolo:1995fq,Pasquali:1998ji} for calculating the flux. 
In Ref.\ \cite{Gondolo:1995fq} it was shown
that this approximate solution agrees with a numerical
Monte Carlo solution of the same equations.

The flux is calculated as a function of the slant depth $X$, which is a measure
of the amount of atmosphere traversed by the particle. It is defined as the
integral of the atmospheric density 
along its path through the atmosphere: 
\be
X(\ell,\theta) = \int_\ell^\infty d\ell' \rho(h(\ell',\theta)),
\ee 
where $h(\ell,\theta)$ is the height at distance from the ground $\ell$ and 
zenith angle $\theta$. 
A reasonable model for our purposes is an exponential atmosphere with \cite{Gondolo:1995fq}
\be
\label{eq:rhoh}
\rho(h) = \rho_0\exp(-h/h_0),
\ee
with $h_0= 6.4$ km and $\rho_0= 2.03\times 10^{-3}$ g/cm$^3$. The vertical
depth of the atmosphere is $X\simeq 1300$ g/cm$^2$, while the horizontal depth is
$X\simeq 36,000$ g/cm$^2$. 
We shall mostly be concerned with the vertical flux,
$\theta=0$, as the conventional flux is the smallest in the vertical direction.
We will, however, show predictions for the flux in the horizontal direction as
well.

The general form of the cascade
equation for the flux $\phi_j=\phi_j(X,E)$ of particle species $j$ at energy
$E$ and slant depth $X$ is
\be
\D{\phi_j}{X} = -\frac{\phi_j}{\lambda_j} 
-\frac{\phi_j}{\lambda_j^\text{dec}} + \sum_k S( k \to j ),
\ee
where $\lambda_j$ is the interaction length,  $\lambda_j^\text{dec}$ is the
decay length, and $S( k \to j )$ is the regeneration function, given by
\be
S( k \to j ) = \int_E^\infty dE' \f{\phi_k(E')}{\lambda_k(E')} 
\D{n(k \to j ;E',E)}{E}. 
\ee

For the case of production,
\be
\D{n(k  \to j ;E_k,E_j)}{E_j} = \f{1}{\sigma_{kA}(E_k)}
\D{\sigma(k A \to j Y,E_k,E_j)}{E_j}
\ee
is the distribution of secondary hadrons and $\sigma_{kA}$ is the total
inelastic cross section for $kA$ collisions. For the case of decays,
\be
\D{n(k  \to j ;E_k,E_j)}{E_j} = \f{1}{\Gamma_{k}}
\D{\Gamma(k\to j Y,E_j)}{E_j}.
\ee
The nucleon, meson, and lepton fluxes are described by the equations
\begin{align}
\D{\phi_N}{X} &= -\frac{\phi_N}{\lambda_N} + S( N A \to N Y ) 
\label{nucleonflux} \\
\D{\phi_M}{X} &= S( N A \to M Y ) -\frac{\phi_M}{\rho d_M(E)} \nn\\ 
&-\frac{\phi_M}{\lambda_M} + S( M A \to M Y ) 
\label{mesonflux} \\
\D{\phi_\ell}{X} &= \sum_M S( M \to \ell Y ) 
\label{leptonflux} 
\end{align}
where $\ell = \mu,\nu_\mu,\nu_e$ and the mesons include unstable baryons:\ for
prompt fluxes from charm $M= D^\pm$, $D^0$, $\bar D^0$, $D_s^\pm$,
$\Lambda_c^\pm$. In \Eref{mesonflux} $d_M=c\beta\gamma\tau$ is the decay
length.

The analytic solution relies on the approximate
factorization of the fluxes into energy- and $X$-dependent parts. 
For the meson flux: 
\be
\D{\phi_M}{X} = - \frac{\phi_M}{\rho d_M} - \frac{\phi_M}{\lambda_M} 
+Z_{MM} \frac{\phi_M}{\lambda_M} + Z_{NM}  \frac{\phi_N}{\lambda_N} 
\label{phiM}
\ee
with
\be
Z_{kh}=\int_E^\infty dE' 
\frac{\phi_k(E',X,\theta)}{\phi_k(E,X,\theta)}\frac{\lambda_k(E)}{\lambda_k(E')}
\D{n(k A \to h Y;E',E)}{E}.
\ee
We now make the standard assumption that
$\phi_k(E,X,\theta)=E^{-\beta_k}\phi_k(X,\theta)$,
so that if the energy spectrum falls as $E^{-\gamma-1}$, we have
\be
Z_{kh}=\int_E^\infty dE' 
\left(\frac{E'}{E}\right)^{-\gamma-1} \frac{\lambda_k(E)}{\lambda_k(E')}
\D{n(k A \to h Y;E',E)}{E}.
\ee
\Eref{nucleonflux} for the nucleon flux then has the solution
\be
\phi_N(X,E)=\phi(E) e^{-X/\Lambda_N(E)},
\ee
where $\phi(E) \equiv \phi(0,E)$ is the primary flux of nucleons on
the atmosphere and $\Lambda_N(E)$ is the nucleon attenuation length, defined as
\be
\Lambda_N(E) =  \frac{\lambda_N(E)}{1-Z_{NN}(E)},
\ee
where $\lambda_N(E)$ is the interaction length of nucleons in the atmosphere. It
is given by
\be
\lambda_N(E) = \frac{A}{N_0 \sigma_{pA}(E)},
\ee
where $A=14.5$ is the average atomic number of air, $N_0$ is Avogadro's number, and
$\sigma_{pA}$ is the total nucleon--air cross section. We take the
parametrization from \cite{Bugaev:1998bi} for this cross section, and the Monte
Carlo result from \cite{Gondolo:1995fq} for $Z_{NN}(E)$.

The meson fluxes are expressed in terms of the nucleon flux by
solving the cascade equations separately at low and high energies, where the
interaction and regeneration terms and the decay terms, respectively, can be
neglected. For the high energy flux we need the attenuation lengths of charmed
hadrons in the atmosphere, which we replace by the corresponding quantities for
$K$-mesons. These are approximated by
\be
\Lambda_M(E) = \frac{A}{N_0 \sigma_{pA}(E)}
\frac{\sigma_{pp}(E)}{\sigma_{Kp}(E)} \frac{1}{1-Z_{KK}(E)}.
\ee
As for nucleons, we take $Z_{KK}$ from \cite{Gondolo:1995fq} and $\sigma_{pA}$
from \cite{Bugaev:1998bi}. The cross sections $\sigma_{pp}$ and $\sigma_{Kp}$
are taken from \cite{Yao:2006px}.

The final step is to obtain the lepton fluxes at high and low energies from
\Eref{leptonflux} and the obtained meson fluxes, and interpolating between them
for intermediate energy. This calculation is done in the limit $X\to \infty$.
The $Z$-moments for the three-body decay modes $M\to \ell Y$ are calculated
using expressions in Refs.\ \cite{Lipari:1993hd,Bugaev:1998bi}, and the lepton flux at
intermediate energies is obtained by interpolating between the high- and
low-energy solutions. In each of these regimes the meson fluxes are described by
power laws $\phi_M(E)\propto E^{-\beta}$ where $\beta=\gamma$ in the low-energy 
regime and $\beta=\gamma+1$ in the high energy regime, and $\gamma$ is the index
of the primary nucleon flux. The higher power of energy in the high energy flux
is due to the appearance of the gamma factor in the decay length in the
denominator of the meson flux.

The equations for the lepton fluxes then give
\begin{align}
\phi^\text{low}_\ell &= Z_{M\ell,\gamma+1} \frac{Z_{NM}}{1-Z_{NN}}\phi_N(E) \\
\phi^\text{high}_\ell &= Z_{M\ell,\gamma+2} \frac{Z_{NM}}{1-Z_{NN}}
\frac{\ln(\Lambda_M/\Lambda_N)}{1-\Lambda_N/\Lambda_M} \frac{\epsilon_M}{E}
\phi_N(E), 
\end{align}
where $\epsilon_M$, the critical energy for meson $M$, separates the low- and
high-energy regions, where attenuation is dominated by decay and interaction. It
depends on zenith angle, and is for the specific model of the atmosphere we use 
given by
\be
\epsilon_M(\theta) = \frac{m_M c^2 h_0}{c\tau_M} f(\theta),
\ee
where $h_0=6.4$~km is a scale parameter for the isothermal height dependence of
the atmospheric density~\cite{Gondolo:1995fq}.
For relatively small angles, $f(\theta)=1/\cos \theta$, but for angles near
horizontal, the angular dependence is more complicated. To compute the 
horizontal flux,
we follow the approach of Ref.\ \cite{Lipari:1993hd}, 
leading to the replacement 
$\theta=90^\circ\to \theta^*=84.45^\circ$. 

Further details of this procedure to solve the cascade equations semi-analytically 
are given, e.g., in Refs.\ \cite{Gondolo:1995fq,Pasquali:1998ji}. 
Our treatment here adds the fragmentation of the charm quarks into charmed
hadrons, meaning that we must compute separately the moments $Z_{ph}$ for each
hadron $M$, including fragmentation functions in the calculation of the cross
section. 
When fragmentation is neglected, we have the simple relation
$Z_{ph} = f_h Z_{pc}$,
where $f_h$ is the fragmentation fraction.

\section{Results and discussion}
\begin{figure} 
\begin{center} 
\epsfig{file=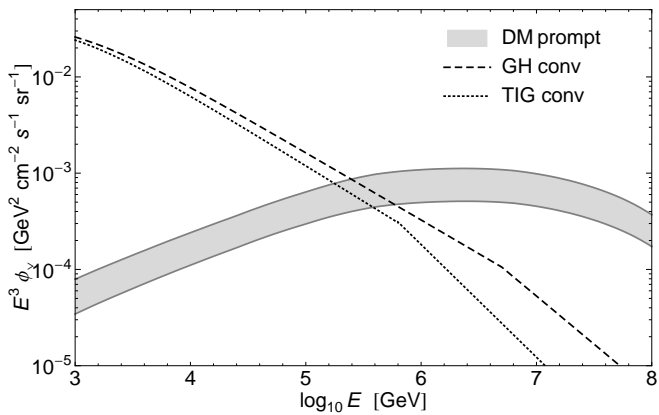, width=\columnwidth}
\caption{Prompt and conventional $\nu_\mu + \bar\nu_\mu$ fluxes in the vertical
direction. The shaded band
is the theoretical uncertainty band for the prompt flux calculated in this
paper with the dipole model. 
The dashed line shows the conventional flux from Gaisser and Honda (GH)
\cite{gaisserhonda} and the dotted line is the conventional flux calculated 
in Ref.~\protect\cite{Gondolo:1995fq} (TIG).
\label{fig:errorband}}  
\end{center}
\end{figure}

Our result for the vertical muon neutrino plus antineutrino flux from atmospheric charm is shown in 
Figure~\ref{fig:errorband}, which shows the theoretical uncertainty band for the
DM calculation, estimated as described above. For comparison the conventional
neutrino fluxes \cite{gaisserhonda,Gondolo:1995fq} 
from $\pi$- and $K$-decays are also shown.
We find that the vertical prompt muon neutrino flux becomes dominant 
over the conventional neutrino flux at energies between $10^5$ GeV and 
$10^{5.5}$ GeV. 

The theoretical uncertainty due to choices of gluon distribution, 
charm quark mass, factorization scale, and other parameters 
in the dipole model result in the range of fluxes represented by the shaded 
area in Figure~\ref{fig:errorband}.  
The shape of the prompt neutrinos is only weakly dependent 
on the choice of parameters, but the overall normalization 
could vary by up to a factor of two in this model for charm production.

\begin{figure} 
\begin{center} 
\epsfig{file=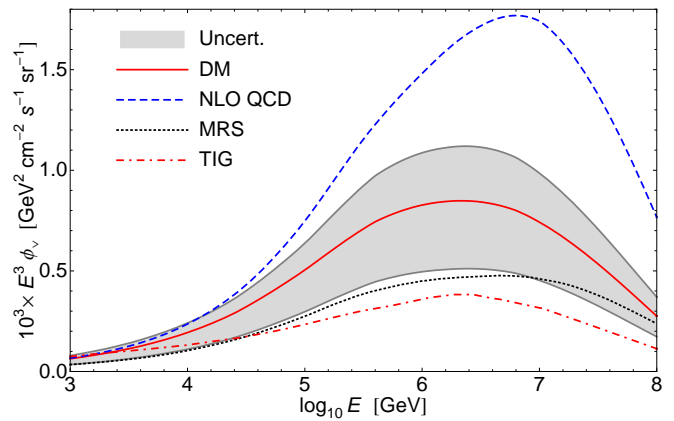, width=\columnwidth}
\caption{Prompt muon neutrino fluxes obtained in perturbative QCD.  
The shaded area represents the
theoretical uncertainty in the prompt neutrino flux evaluated in this paper, 
and the solid line in the band is our standard result. 
The dashed curve is the NLO 
perturbative QCD calculation of Ref.
\protect\cite{Pasquali:1998ji} (PRS), modified here
to include fragmentation, the dotted  curve is the saturation model result
of Ref.~\protect\cite{Martin:2003us} (MRS), and the dash-dotted curve is 
the LO perturbative QCD calculation of Ref.\ \protect\cite{Gondolo:1995fq} (TIG).
\label{fig:compare_others}}  
\end{center}
\end{figure}

We compare our result to three earlier calculations of the 
prompt neutrino flux:
\begin{enumerate}
\item
Thunman, Ingelman and Gondolo (TIG)~\cite{Gondolo:1995fq}. This was the first  perturbative QCD
calculation and was done at the leading order (LO) in $\alpha_s$. It takes the
fragmentation of charm quarks into account through Monte Carlo simulation
using the Lund string model~\cite{Andersson:1983ia} implemented in the event 
generator Pythia~\cite{pythia}. The small-$x$ PDFs are extrapolated with
e.g., $xG(x,\mu^2)\sim x^{-0.08}$.
\item
Pasquali, Reno and Sarcevic (PRS)~\cite{Pasquali:1998ji}. 
This result uses the next-to-leading order (NLO) QCD result of~\cite{NLO}
with power law extrapolations of the small-$x$ PDFs.
The PRS evaluation 
does not take fragmentation into account. We have therefore carried out
a simplified version of this calculation, taking fragmentation into account in
the same way as we did for the DM calculation:\ we 
compute the charmed hadron cross section in leading order QCD using
KK fragmentation functions~\cite{KniehlKramer}, and multiply with a K-factor 
$K=\sigma(NLO)/\sigma(LO) \approx 2$.
This reproduces the full NLO calculation of Ref.~\cite{Pasquali:1998ji} 
at the parton level to an adequate accuracy.
\item
Martin, Ryskin and Sta\'sto (MRS)~\cite{Martin:2003us}. 
This calculation takes
fragmentation into account by assigning the neutrino a fixed fraction of the momentum of the mother meson, and is done using the
saturation model of
Golec-Biernat and W\"ustoff~\cite{GolecBiernat:1998js} described above.
\end{enumerate}

We show the results from these other evaluations 
of the vertical muon neutrino plus antineutrino
flux together with our uncertainty band in 
Figure~\ref{fig:compare_others}. 
The theoretical uncertainties in the standard NLO QCD calculation of the
charm cross section are the choice of the
renormalization and factorization scales, the charm mass, and the small $x$ behavior
of the gluon distribution~\cite{Vogt:2007aw}. The impact of some of these uncertainties on the
neutrino flux has been studied in Ref.\ \cite{graciela}.

The MRS curve in Fig. \ref{fig:compare_others} is at the lower border
of our DM uncertainty band. There is approximately a factor of two
between the MRS and the central DM results, coming from the different
parameterizations of $\sigma_d$.  The enhancement is also seen in
calculations of photoproduction of heavy quarks
\cite{Goncalves:2004dn} comparing the GBW model and the improved DM
model of Eq. \ref{sigmadip1}.
The DM cross section for charm pair production in $pp$ collisions lies
within the uncertainty band of Ref. \cite{Vogt:2007aw}.

\begin{figure}[t]  
\begin{center} 
\epsfig{file=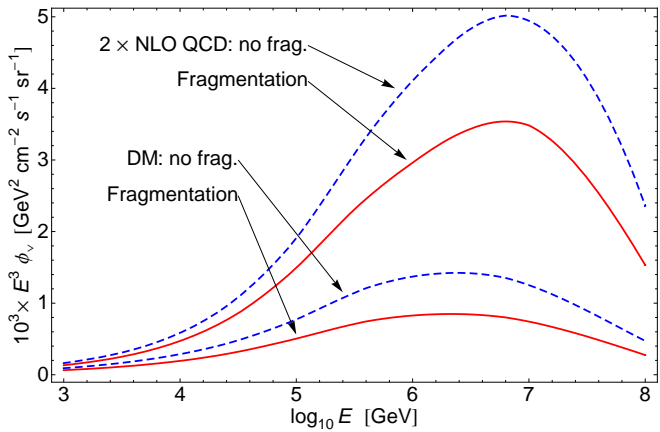, width=\columnwidth}
\caption{The effect of fragmentation on predicted $\nu_\mu + \bar\nu_\mu$
fluxes. The solid lines are DM and NLO QCD (the latter multiplied
by two to separate the lines) with Kniehl--Kramer
fragmentation. The dashed lines are without fragmentation.
\label{fig:fragmentation}}  
\end{center}
\end{figure}

The effect of quark fragmentation on the neutrino fluxes is
rather large because fragmentation reduces the energy 
of the charmed hadrons. For a given hadron energy, fragmentation
effects require  higher energy cosmic rays in the steeply falling
cosmic ray flux. 
In Figure~\ref{fig:fragmentation} we show the effect of including the KK
fragmentation functions on both the NLO QCD and DM results. 
The NLO results are multiplied by a factor of two so that
they can be distinguished easily from the DM results.
The fragmentation
reduces the flux by between 60\% and 70\%, and thus it is an important effect to take
into account. The Peterson fragmentation function 
results differ by approximately 10\% from the results shown in 
Figure~\ref{fig:fragmentation}. We include
the uncertainty in fragmentation model in our estimate of the theoretical 
uncertainty; however, we do not consider the result without fragmentation in our
uncertainty estimate.

\begin{figure}[tb] 
\begin{center} 
\epsfig{file=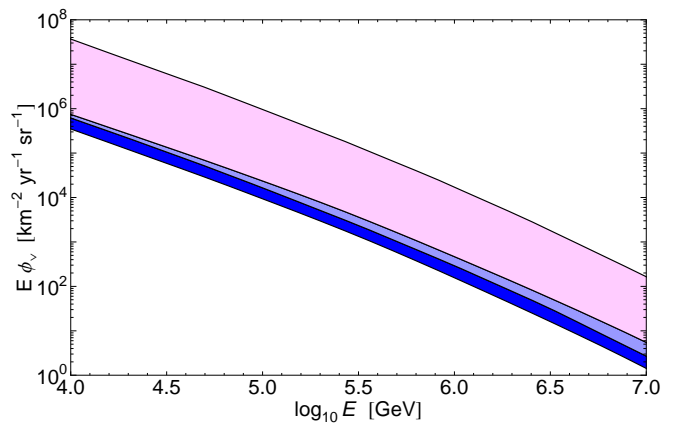, width=\columnwidth}
\caption{Theoretical uncertainty band (blue, dark band), 
 compared to the uncertainty band from Ref.\ \cite{ando}  
(magenta, light band) with the 
 overlapping region shown as the middle (light blue) band. 
The flux is scaled by $E$, in units of 1/km$^2$yr\,sr.
\label{fig:compare_ABPR}}  
\end{center}
\end{figure}

Other perturbative QCD calculations are unlikely to give a much 
larger prompt neutrino flux than our upper limit, if saturation is indeed 
important.
The theoretical expectation is that saturation is important at
scales comparable to $\mu\sim m_c$ for small $x$ values. 
If it would turn out that
saturation \emph{does not} occur at the relevant energy scales, the flux is 
still not expected to be much larger than the PRS result. In 
Figure~\ref{fig:compare_ABPR},
we therefore show a comparison of the uncertainty (blue, dark band) 
compared to the proposed uncertainty range
from Ref.~\cite{ando} (magenta, light band) with their overlapping 
region (middle, light blue band).  In this plot we take the NLO QCD result
as the upper theoretical limit. This gives a larger upper limit than in the 
earlier plots, which show only the uncertainty in the dipole model result. 
We stress that, since saturation is expected to be important on theoretical 
grounds, the uncertainty band in Figure \ref{fig:errorband} is our main result. 

In order to obtain a flux as large as the upper line 
in the  uncertainty band of Figure~\ref{fig:compare_ABPR}, we must multiply the
upper uncertainty line of our DM result
by a factor of 50. A cross section a factor of 50 times
larger than the DM evaluation in proton-proton 
scattering would be incompatible with existing cross section measurements,
as illustrated for example by Figure~6 of Ref.~\cite{Goncalves:2006ch}, which
compares the DM result for charm production to fixed-target experimental data.

Measurable stau production
rates from prompt atmospheric neutrinos as
proposed in Ref.~\cite{ando} would require the highest
fluxes in the lighter band.
Our evaluation of the prompt neutrino flux indicates that the
upper limit of Ref.\ \cite{ando} is unrealistically large.
The prompt neutrino flux
is unlikely
to be large enough for studying stau production from neutrino interactions with
Earth and the subsequent detection in neutrino telescopes. 

\begin{figure}[t] 
\begin{center} 
\epsfig{file=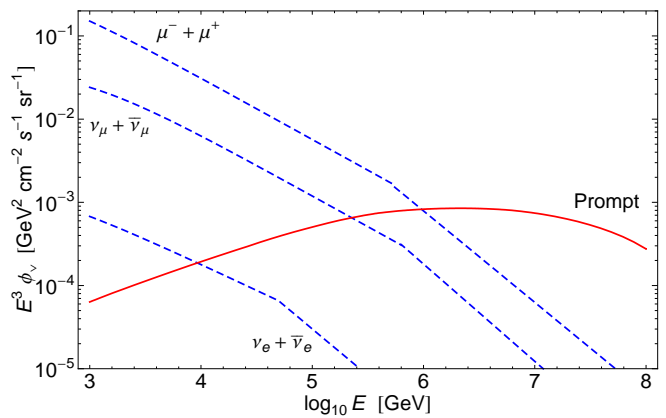, width=\columnwidth}
\caption{Prompt (solid line) and conventional (dashed lines) fluxes
of $\nu_\mu + \bar\nu_\mu$, $\nu_e + \bar\nu_e$, and $\mu^+ + \mu^-$. The
conventional fluxes are from Ref.~\protect\cite{Gondolo:1995fq}. The three prompt 
fluxes are approximately equal and are therefore here represented by the 
$\nu_\mu + \bar\nu_\mu$ flux.
\label{fig:flavors}}  
\end{center}
\end{figure}

The flavor decomposition of an atmospheric neutrino signal may be
an interesting way to explore the prompt contribution.
The prompt neutrino fluxes of $\nu_\mu + \bar\nu_\mu$ and $\nu_e + \bar\nu_e$ 
are identical, since the charmed mesons decay equally likely into electrons or
muons. The prompt 
$\mu^+ + \mu^-$ flux is approximately equal to the neutrino fluxes.
However, this does not hold for the conventional fluxes. Charged pions decay
almost exclusively into muons, so the muon neutrino and muon fluxes are much
larger than the electron neutrino flux. We show the $\nu_\mu + \bar\nu_\mu$ 
prompt flux together with the corresponding vertical conventional fluxes 
of muons, muon neutrinos and electron neutrinos (and their antiparticles)
in
Figure~\ref{fig:flavors}. If experiments would be able to measure electron
neutrino fluxes, the prompt flux will start dominating over the conventional
flux for much lower energy $\sim 10^4$ GeV than for the muon neutrino or muon
fluxes.

\begin{figure}[tb]
\begin{center} 
\epsfig{file=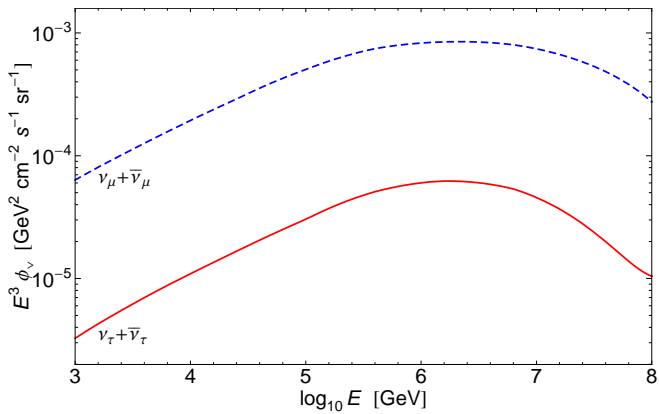, width=\columnwidth}
\caption{Prompt $\nu_\tau + \bar\nu_\tau$ flux (solid line) compared with 
the prompt $\nu_\mu + \bar\nu_\mu$ flux (dashed line).
\label{fig:nutau}}  
\end{center}
\end{figure}

\begin{figure}[t] 
\begin{center} 
\epsfig{file=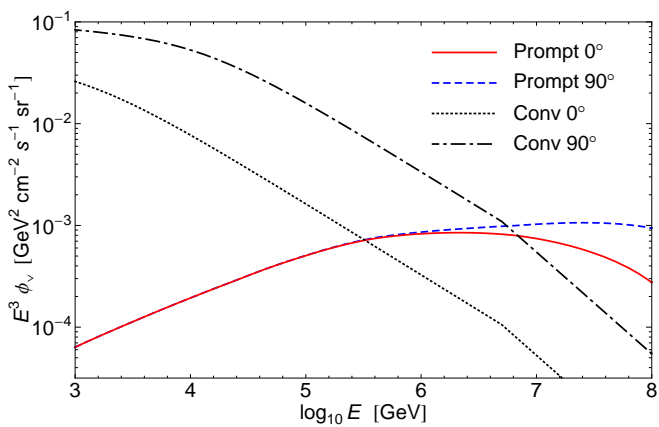, width=\columnwidth}
\caption{Dependence on zenith angle of prompt and conventional 
$\nu_\mu + \bar\nu_\mu$ fluxes. The solid  and dashed lines are DM 
with fragmentation in the vertical and horizontal directions, respectively. The
dotted and dash-dotted lines are the Gaisser--Honda conventional fluxes \cite{gaisserhonda}
in the vertical and horizontal directions.
\label{fig:direction}}  
\end{center}
\end{figure}

We note that the prompt flux of $\nu_\tau + \bar\nu_\tau$ from
charm decays is much smaller than the other 
neutrino flavors \cite{tau}, 
since only the $D_s$ meson decays into $\nu_\tau$. The
$\nu_\tau + \bar\nu_\tau$ 
flux from $D_s$ decays 
is shown in Figure~\ref{fig:nutau} together with the 
prompt $\nu_\mu + \bar\nu_\mu$ flux. We have not included the contribution from
$B$-meson decays which could give a contribution on the order of 
10--20\%~\cite{Martin:2003us} since $B$-meson decays to $\nu_\tau$ plus
tau are kinematically allowed.

The vertical direction is the optimal direction for studying
the prompt fluxes. In Figure~\ref{fig:direction} we show the prompt and
conventional $\nu_\mu + \bar\nu_\mu$ fluxes in the vertical and horizontal
directions. In the horizontal direction the prompt flux does not become larger
than the conventional flux until very large energies $\sim 10^7$ GeV, where the actual
number of neutrinos is quite low.

In summary, we have computed prompt neutrino and muon
fluxes from cosmic ray interactions in the atmosphere that produce
charm pairs. Our evaluation of the fluxes takes 
parton saturation effects into account via the dipole model, a model
with a parametric form guided by QCD and constrained by data. 
We find that saturation effects in the dipole model decrease
the prompt fluxes above $10^5$~GeV.  Our estimate of the theoretical
uncertainty in the predicted fluxes in the dipole model is on the order
of a factor of two. In comparison to other QCD or dipole model
evaluations of the prompt flux,
the range of predictions is approximately a factor of 6. Future 
measurements of the high energy neutrino flux will provide interesting
constraints on QCD-based
evaluations of the prompt flux of neutrinos, however, the prompt
neutrino flux is 
unlikely to be large enough to probe non-standard model interactions.

\begin{acknowledgments}
We would like to thank Julia Becker, Magno Machado and Teresa Montaruli 
for helpful discussions. We are also grateful to Magno Machado for providing us with 
details of the
results of \cite{Goncalves:2006ch}.  The Feynman diagrams in
Figures~\ref{fig:1} and  \ref{fig:2} where drawn using JaxoDraw~\cite{jaxodraw}.
This research was supported in part by US Department of Energy contracts
DE-FG02-91ER40664, DE-FG02-04ER41319 and DE-FG02-04ER41298.
\end{acknowledgments}


\end{document}